\documentclass[12pt]{article}
\usepackage[dvips]{epsfig}
\begin{document}
\begin{flushright}
OHSTPY-HEP-T-98-029 \\
hep-th/9812036
\end{flushright}
\vspace{20mm}
\begin{center}
{\LARGE Hyperbolic Numbers and the Dirac Spinor}
\\
\vspace{20mm}
{\bf Francesco Antonuccio \\}
\vspace{4mm}
Department of Physics,\\ The Ohio State University,\\ Columbus, OH 43210, USA\\
\vspace{4mm}
\end{center}
\vspace{10mm}

\begin{abstract}
A representation of the Lorentz group
is given in terms of $4 \times 4$ matrices defined over
the hyperbolic number system. The transformation properties of
the corresponding four component spinor are studied, and 
shown to be equivalent to the transformation properties of
the complex Dirac spinor. 
As an application, we show that there exists
an algebra of automorphisms of the complex
Dirac spinor that leaves the transformation
properties of its eight real components 
invariant under any given Lorentz transformation.   
Interestingly, the representation of the Lorentz
algebra presented here is naturally embedded in
the Lie algebra of a group isomorphic to SO(3,3;${\bf R}$) instead
of the conformal group SO(2,4;${\bf R}$).

\end{abstract}
\newpage

\baselineskip .25in

\section{Introduction}
This article is motivated by the simple observation that the 
transformation properties of the eight real components
of a complex Dirac spinor under a Lorentz
transformation may be alternatively 
formulated without 
any explicit reference
to complex-valued quantities.
This is accomplished by constructing a representation of the 
Lorentz group in terms of $4 \times 4$ matrices
defined over the hyperbolic number system 
\cite{cap}--\cite{yaglom}.
After studying how this new representation is related to
the familiar complex one, we establish
an automorphism symmetry of the complex Dirac spinor.
We also discuss natural embeddings of this new representation
into a maximal Lie algebra, which turns out to be 
isomorphic to the algebra of generators of SO(3,3;${\bf R}$),
and thus distinct from the conformal group SO(2,4;${\bf R}$).

To begin, we revisit the familiar Lie algebra of the
Lorentz group O(1,3;${\bf R}$). 
\section{The Lorentz Algebra}
\subsection{A Complex Representation}
Under Lorentz transformations, the complex Dirac 4-spinor $\Psi_{\bf C}$
transforms as follows \cite{ryder}:
\begin{equation}
\Psi_{\bf C} \rightarrow \left(
\begin{array}{cc}
       e^{\frac{{\rm i}}{2} 
   \mbox{${\bf \sigma \cdot}$} 
          ( \mbox{${\bf \theta}$} - {\rm i}
           \mbox{${\bf \phi}$})}
            & 0 \\
       0 &  e^{\frac{{\rm i}}{2} 
   \mbox{${\bf \sigma \cdot}$} 
          ( \mbox{${\bf \theta}$} + {\rm i}
           \mbox{${\bf \phi}$})} 
\end{array}
\right) \cdot \Psi_{\bf C},
\label{lorentz1} 
\end{equation}
where ${\bf \sigma} = (\sigma_{x},\sigma_{y},\sigma_{z})$ represents
the well known Pauli spin matrices:
\begin{equation}
\sigma_x = \left(
\begin{array}{cc}
 0 & 1 \\
 1 & 0
\end{array}
\right), \hspace{5mm}
\sigma_y = \left(
\begin{array}{cc}
 0 & -{\rm i} \\
 {\rm i} & 0
\end{array}
\right), \hspace{5mm}
\sigma_z = \left(
\begin{array}{cc}
 1 & 0 \\
 0 & -1
\end{array}
\right).
\end{equation}
The three real parameters ${\bf \theta} = (\theta_1,\theta_2,\theta_3)$
correspond to the generators for spatial rotations,
while ${\bf \phi} = (\phi_1,\phi_2,\phi_3)$ represents 
Lorentz boosts along each of the coordinate axes.
There are thus six real numbers parameterizing a given element
in the Lorentz group.

Let us now introduce the six matrices $E_i$ and $F_i$, $i=1,2,3$,
by writing
\begin{equation}
\begin{array}{ccc}
E_1 = \frac{1}{2} \left( 
\begin{array}{cc}
\sigma_x & 0 \\
0 & -\sigma_x
\end{array}
\right)
& 
E_2 = -\frac{{\rm i}}{2} \left( 
\begin{array}{cc}
\sigma_y & 0 \\
0 & \sigma_y
\end{array}
\right)
 & E_3 = \frac{1}{2} \left( 
\begin{array}{cc}
\sigma_z & 0 \\
0 & -\sigma_z
\end{array}
\right)  \\
F_1 = \frac{{\rm i}}{2} \left( 
\begin{array}{cc}
\sigma_x & 0 \\
0 & \sigma_x
\end{array}
\right)
 & 
F_2 = \frac{1}{2} \left( 
\begin{array}{cc}
\sigma_y & 0 \\
0 & -\sigma_y
\end{array}
\right)
 & 
F_3 = \frac{{\rm i}}{2} \left( 
\begin{array}{cc}
\sigma_z & 0 \\
0 & \sigma_z
\end{array}
\right).
\end{array}
\label{ef}
\end{equation}
Then the Lorentz transformation (\ref{lorentz1}) may be written as follows:
\begin{equation}
\Psi_{\bf C} \rightarrow \exp{(\phi_1 E_1 - \theta_2 E_2 + \phi_3 E_3 + 
\theta_1 F_1 + \phi_2 F_2 + \theta_3 F_3)} \cdot  \Psi_{\bf C}.
\label{transformD}
\end{equation}
We remark that the transformation (\ref{transformD}) has the form
$\Psi_{\bf C} \rightarrow U \cdot \Psi_{\bf C}$, where $U$ may
be thought of as an element of the (fifteen dimensional)
conformal group SU(2,2;${\bf C}$). The Lorentz symmetry
is therefore a six dimensional subgroup of the conformal group. 

At this point, it is sufficient to note that the matrices
$E_i$ and $F_i$ defined in (\ref{ef}) satisfy the following
commutation relations:
\begin{equation}
\begin{array}{llll}
[E_1,E_2] = E_3 & [F_1,F_2] = -E_3 & [E_1,F_2] = F_3 & [F_1,E_2]=F_3 \\
\mbox{}[E_2,E_3] = E_1 & [F_2,F_3] = -E_1 & [E_2,F_3] = F_1 & [F_2,E_3]=F_1 \\
\mbox{}[E_3,E_1] = -E_2 & [F_3,F_1] = E_2 & [E_3,F_1] = -F_2 & [F_3,E_1]=-F_2
\end{array}
\label{comm}
\end{equation}
All other commutators vanish. Abstractly, these relations define
the Lie algebra of the Lorentz group O(1,3;${\bf R}$), and the matrices
$E_i$ and $F_i$ defined by (\ref{ef}) correspond to a complex
representation of this algebra. 
\subsection{A Hyperbolic Representation}
Our goal in this section is to present an explicit
representation of the Lorentz algebra (\ref{comm}) in terms 
of $4 \times 4$ matrices
defined over the hyperbolic number system.
This number system will be briefly discussed next.

\subsubsection{The Hyperbolic Number System}
We consider numbers of the form
\begin{equation}
x+ {\rm j}y,
\end{equation}
where $x$ and $y$ are real numbers, and ${\rm j}$ is a commuting
element satisfying the relation
\begin{equation}
 {\rm j}^2 = 1.
\end{equation}
The number system generated by this simple algebra has
a long history \cite{cap}--\cite{yaglom}, and is known as
the `hyperbolic number system'. The
symbol ${\bf D}$ will be used to denote the hyperbolic number
system, where `D' stands for `double' \cite{yaglom}.  

In this article, we exploit very basic arithmetical properties
of this algebra. For example,
addition, subtraction, and multiplication are defined in the obvious way:
\begin{eqnarray}
 (x_1+{\rm j}y_1) \pm (x_2+{\rm j}y_2) & = & 
 (x_1 \pm x_2) + {\rm j}(y_1 \pm y_2), \\
 (x_1+{\rm j}y_1) \cdot (x_2+{\rm j}y_2) & = & 
 (x_1 x_2 + y_1 y_2) + {\rm j} (x_1 y_2 + y_1 x_2).
\end{eqnarray}
Moreover, given any hyperbolic number $w=x+{\rm j}y$, we define the 
`${\bf D}$-conjugate of $w$', written ${\overline w}$, to be 
\begin{equation}
 {\overline w} = x - {\rm j}y.
\end{equation}
It is easy to check the following; for any $w_1,w_2 \in {\bf D}$,
we have 
\begin{eqnarray}
{\overline{w_1+w_2}} & = & {\overline w_1} + {\overline w_2}, \\
{\overline{w_1\cdot w_2}} & = & {\overline w_1} \cdot {\overline w_2}.
\end{eqnarray}
We also have the identity
\begin{equation}
 {\overline w} \cdot w = x^2 - y^2
\end{equation}
for any hyperbolic number $w=x+{\rm j}y$. Thus ${\overline w} \cdot w$
is always real, although unlike
the complex number system, it may take negative values.

At this point, it is convenient to define the `modulus squared'
of $w$, written $|w|^2$, as  
\begin{equation}
       |w|^2 = {\overline w} \cdot w.
\end{equation}
A nice consequence of these definitions is that for
any hyperbolic numbers $w_1,w_2 \in {\bf D}$, we have
\begin{equation}
 |w_1 \cdot w_2|^2 = |w_1|^2 \cdot |w_2|^2.
\end{equation}
Now observe that if $|w|^2$ doesn't vanish, the quantity
\begin{equation}
w^{-1} = \frac{1}{|w|^2} \cdot {\overline{w}}
\end{equation}
is a well-defined unique inverse for $w$. So $w\in {\bf D}$ fails to
have an inverse if and only if $|w|^2 = x^2-y^2 =0$.
The hyperbolic number system is therefore a non-division algebra.
\subsubsection{The Hyperbolic Unitary Groups}
Suppose $H$ is an $n \times n$ matrix defined over ${\bf D}$.
Then $H^{\dagger}$ will denote the $n\times n$ matrix
which is obtained by transposing $H$, and then conjugating 
each of the entries: $H^{\dagger} = \overline{H}^{T}$.
We say $H$ is Hermitian with respect to ${\bf D}$ if
$H^{\dagger} = H$, and anti-Hermitian if
$H^{\dagger} = -H$. 

Note that if $H$ is an $n \times n$ Hermitian matrix over ${\bf D}$,
then $U = e^{{\rm j}H}$ has the property 
\begin{equation}
           U^{\dagger}\cdot U = U \cdot U^{\dagger} = 1.
\label{un}
\end{equation}  
The set of all $n \times n$ matrices over ${\bf D}$ satisfying
the constraint (\ref{un}) forms a group, which we will
denote as U$(n,{\bf D})$, and call the `unitary group of
$n \times n$ matrices over ${\bf D}$', or 
`hyperbolic unitary group'.
The `special unitary' subgroup SU($n,{\bf D}$) will be defined as all 
elements $U \in$U($n,{\bf D}$) satisfying the additional constraint
\begin{equation}
\det{U} = 1.
\end{equation}
Note that the hyperbolic unitary groups we have defined above 
may be isomorphic to well known non-compact groups 
that are usually defined over the complex number field. 
For example, the special unitary hyperbolic
group SU$(2,{\bf D})$ is isomorphic to the 
complex group SU(1,1;${\bf C}$) by virtue of the 
identification\footnote{To show that any $2 \times 2$ matrix 
$U \in SU(2,{\bf D})$ has the form given in eqn (\ref{iso2DD}),
we use the facts $U^{\dagger} = U^{-1}$, and $\det U =1$.}  
\begin{equation}
\left(
\begin{array}{cc}
a_1 + {\rm i}a_2 & b_1 + {\rm i}b_2 \\
b_1 - {\rm i}b_2 & a_1 - {\rm i} a_2
\end{array} 
\right) \hspace{4mm}
\leftrightarrow \hspace{4mm}
\left(
\begin{array}{cc}
a_1 + {\rm j}b_1 & -a_2 + {\rm j}b_2 \\
a_2 + {\rm j}b_2 & a_1 - {\rm j} b_1 
\end{array}
\right),
\label{iso2DD}
\end{equation}
where the four real parameters $a_1,a_2,b_1$ and $b_2$ satisfy
the constraint $a_1^2+a_2^2-b_1^2-b_2^2=1$.

Of course, one also has the isomorphism SU$(2,{\bf D}) \equiv SL(2;{\bf R})$
given by the group isomorphism
\begin{equation}
\left(
\begin{array}{cc}
a_1 + {\rm j}b_1 & -a_2 + {\rm j}b_2 \\
a_2 + {\rm j}b_2 & a_1 - {\rm j} b_1 
\end{array} 
\right) \hspace{4mm}
\leftrightarrow \hspace{4mm}
\left(
\begin{array}{cc}
a_1 + b_1 & -a_2 + b_2 \\
a_2 + b_2 & a_1 - b_1 
\end{array}
\right),
\label{iso2D}
\end{equation}
where the real parameters $a_1,a_2,b_1$ and $b_2$ satisfy
the constraint $a_1^2+a_2^2-b_1^2-b_2^2=1$ as before.
Note that this correspondence was obtained by mapping the variable
${\rm j}$ to $+1$. Alternatively, we could have constructed
an alternative isomorphism by mapping ${\rm j}$ to $-1$.

Actually, this example suggests that we might be able
to identify the special unitary groups SU($n$;${\bf D}$) 
with the special linear groups
SL($n$;${\bf R}$). An isomorphism was established for $n=2$,
but what can we say about $n > 2$? One approach is to consider
what happens near the identity. In this case, one may construct 
the Lie algebra for SU($n$;${\bf D}$), which is generated by
$n^2 -1$ traceless anti-Hermitian $n \times n$ 
matrices over ${\bf D}$. 
Any element sufficiently close to the
identity is therefore obtained by exponentiating
a unique real linear combination of these generators.
We then map such elements into  SL(n;${\bf R}$)
by mapping the variable ${\rm j}$ to $+1$. The generators 
are now real, traceless $n \times n$ matrices, and so form
the basis of the Lie algebra for SL(n;${\bf R}$). Thus, the groups
SU($n$;${\bf D}$) and SL($n$;${\bf R}$) possess isomorphic
Lie algebras. 

\subsubsection{A Hyperbolic Representation}
As promised, we will give an explicit representation of
the Lorentz algebra (\ref{comm}) in terms of matrices
defined over ${\bf D}$. First, we define three $2 \times 2$ matrices
${\bf \tau} = (\tau_1,\tau_2,\tau_3)$ by writing
\begin{equation}
\tau_1 = \left(
\begin{array}{cc}
 0 & 1 \\
1 & 0 
\end{array}
\right),
\hspace{5mm}
\tau_2 = \left(
\begin{array}{cc}
 0 & -{\rm j} \\
{\rm j} & 0 
\end{array}
\right),
\hspace{5mm}
\tau_3 = \left(
\begin{array}{cc}
 1 & 0 \\
 0 & -1 
\end{array}
\right).
\hspace{5mm}
\end{equation}
These matrices satisfy the following commutation relations:
\begin{equation}
[\tau_1,\tau_2]=2{\rm j} \tau_3 \hspace{6mm} 
[\tau_2,\tau_3]=2{\rm j} \tau_1 \hspace{6mm}
[\tau_3,\tau_1]=-2{\rm j} \tau_2   
\end{equation}
Now define the matrices ${\tilde E}_i$ and ${\tilde F}_i$, 
$i=1,2,3,$ by setting
\begin{equation}
 {\tilde E}_i = \frac{{\rm j}}{2}\left(
\begin{array}{cc}
 \tau_i & 0 \\
 0 & \tau_i
\end{array}
\right),
\hspace{6mm}
 {\tilde F}_i = \frac{1}{2}\left(
\begin{array}{cc}
 0 & \tau_i  \\
 -\tau_i & 0
\end{array}
\right), \hspace{5mm} i=1,2,3.
\label{semirep}
\end{equation}
The $4 \times 4$  matrices ${\tilde E}_i$ and ${\tilde F}_i$
defined above are anti-Hermitian with respect to ${\bf D}$,
and satisfy the Lorentz algebra (\ref{comm})
after making the substitutions $E_i \rightarrow {\tilde E}_i$
and $F_i \rightarrow {\tilde F}_i$, $i=1,2,3$. We may therefore
introduce a 4-component `hyperbolic' spinor 
$\Psi_{\bf D} \in {\bf D}^4$
transforming as follows under Lorentz transformations:
\begin{equation}
 \Psi_{\bf D} \rightarrow 
\exp{(\phi_1 {\tilde E}_1 - \theta_2 {\tilde E}_2 + 
\phi_3 {\tilde E}_3 + 
\theta_1 {\tilde F}_1 + \phi_2 {\tilde F}_2 + \theta_3 
{\tilde F}_3)} \cdot  \Psi_{\bf D},
\label{transformII}
\end{equation}
which is evidently the analogue of transformation 
(\ref{transformD}). Note that the transformation (\ref{transformII})
has the form $\Psi_{\bf D} \rightarrow U \cdot \Psi_{\bf D}$,
where $U \in$ SU($4,{\bf D}$), since the generators 
${\tilde E}_i$ and ${\tilde F}_i$
are traceless and anti-Hermitian with respect to ${\bf D}$. 
Thus the Lorentz group is a {\em subgroup} of the hyperbolic
special unitary group SU($4,{\bf D}$). 

In the next section, we discuss a relation between the
complex Dirac spinor $\Psi_{\bf C}$, and the 4-component
hyperbolic spinor $\Psi_{\bf D}$ defined above. 

\section{Equivalences between Spinor Transformations}
\subsection{An Equivalence}
\label{isomorph}
Consider an {\em infinitesimal} Lorentz transformation of the
complex Dirac spinor,
\begin{equation}
\Psi_{\bf C} \rightarrow \exp{(\phi_1 E_1 - \theta_2 E_2 + \phi_3 E_3 + 
\theta_1 F_1 + \phi_2 F_2 + \theta_3 F_3)} \cdot  \Psi_{\bf C},
\label{transformsmall}
\end{equation}
where 
\begin{equation}
\Psi_{\bf C} =
\left(
\begin{array}{c}
x_1+{\rm i}y_1 \\
x_2+{\rm i}y_2 \\
x_3+{\rm i}y_3 \\
x_4+{\rm i}y_4
\end{array}
\label{cd}
\right),
\end{equation}
and $E_i,F_i$ are specified by (\ref{ef}).
The eight variables
$x_i$ and $y_i$, $i=1,2,3,4$, are taken to be real.
Now consider the corresponding
infinitesimal Lorentz transformation of the hyperbolic spinor
$\Psi_{\bf D}$,
\begin{equation}
\Psi_{\bf D} \rightarrow 
\exp{(\phi_1 {\tilde E}_1 - \theta_2 {\tilde E}_2 + \phi_3 
{\tilde E}_3 + 
\theta_1 {\tilde F}_1 + \phi_2 {\tilde F}_2 + \theta_3 
{\tilde F}_3)} \cdot  \Psi_{\bf D},
\label{transformsmallII}
\end{equation}
where 
\begin{equation}
\Psi_{\bf D} =
\left(
\begin{array}{c}
a_1+{\rm j}b_1 \\
a_2+{\rm j}b_2 \\
a_3+{\rm j}b_3 \\
a_4+{\rm j}b_4
\end{array}
\right).
\end{equation}
The matrices ${\tilde E}_i,{\tilde F}_i$ are given by
(\ref{semirep}), and
the eight variables $a_i$ and $b_i$, $i=1,2,3,4$, are real-valued.

It is now straightforward to check that the infinitesimal 
transformations (\ref{transformsmall}) and 
(\ref{transformsmallII}) 
induce {\em equivalent} transformations of the eight real
components of the corresponding spinors ($\Psi_{\bf C}$ and
$\Psi_{\bf D}$)
if we make the following identifications\footnote{The factor of
$1/\sqrt{2}$ is arbitrary, and introduced for later convenience.}:
\begin{equation}
\begin{array}{cccc}
a_1 \leftrightarrow \frac{1}{\sqrt{2}}(y_1+y_3) \hspace{4mm} & 
a_2 \leftrightarrow \frac{1}{\sqrt{2}}(y_2 + y_4) \hspace{4mm}&
a_3 \leftrightarrow \frac{1}{\sqrt{2}}(x_1 + x_3) \hspace{4mm} &
a_4 \leftrightarrow \frac{1}{\sqrt{2}}(x_2 + x_4)  \\
b_1 \leftrightarrow \frac{1}{\sqrt{2}}(y_1-y_3) \hspace{4mm} &
b_2 \leftrightarrow \frac{1}{\sqrt{2}}(y_2-y_4) \hspace{4mm}  &
b_3 \leftrightarrow \frac{1}{\sqrt{2}}(x_1 - x_3) \hspace{4mm}&
b_4 \leftrightarrow \frac{1}{\sqrt{2}}(x_2 - x_4) 
\end{array} 
\end{equation}
In particular, we have the identification
\begin{equation}
(I) \hspace{3mm} \Psi_{\bf C} =
\left(
\begin{array}{c}
x_1+{\rm i}y_1 \\
x_2+{\rm i}y_2 \\
x_3+{\rm i}y_3 \\
x_4+{\rm i}y_4
\end{array}
\right) \hspace{5mm} \leftrightarrow \hspace{5mm}
\Psi_{\bf D} =
\frac{1}{\sqrt{2}}\left(
\begin{array}{c}
(y_1+y_3)  +{\rm j}(y_1-y_3)  \\
(y_2 + y_4) +{\rm j}(y_2-y_4)  \\
(x_1 + x_3) +{\rm j}(x_1 - x_3) \\
(x_2 + x_4)  +{\rm j}(x_2 - x_4)
\end{array}
\right),
\label{iso}
\end{equation}   
which establishes an exact equivalence between a complex
Lorentz transformation [Eqn(\ref{transformD})] 
acting on the Dirac 4-spinor
$\Psi_{\bf C}$, and the corresponding
Lorentz transformation [Eqn(\ref{transformII})]
acting on a hyperbolic 4-spinor $\Psi_{\bf D}$.

It turns out that
the equivalence specified by the identification (\ref{iso})
is not unique. There are additional identifications that
render the complex and hyperbolic Lorentz transformations
equivalent, and we list three more below:
\begin{equation}
(II) \hspace{3mm} \left(
\begin{array}{c}
x_1+{\rm i}y_1 \\
x_2+{\rm i}y_2 \\
x_3+{\rm i}y_3 \\
x_4+{\rm i}y_4
\end{array}
\right) \hspace{3mm} \leftrightarrow \hspace{3mm}
\frac{1}{\sqrt{2}}\left(
\begin{array}{c}
-(y_2+y_4)  +{\rm j}(y_2-y_4)  \\
(y_1 + y_3) -{\rm j}(y_1-y_3)  \\
(x_2 + x_4) -{\rm j}(x_2 - x_4) \\
-(x_1 + x_3)  +{\rm j}(x_1 - x_3)
\end{array}
\right),
\end{equation}
\begin{equation}
(III) \hspace{3mm} 
\left(
\begin{array}{c}
x_1+{\rm i}y_1 \\
x_2+{\rm i}y_2 \\
x_3+{\rm i}y_3 \\
x_4+{\rm i}y_4
\end{array}
\right) \hspace{3mm} \leftrightarrow \hspace{3mm}
\frac{1}{\sqrt{2}}\left(
\begin{array}{c}
-(x_1+x_3)  -{\rm j}(x_1-x_3)  \\
-(x_2 + x_4) -{\rm j}(x_2-x_4)  \\
(y_1 + y_3)  +{\rm j}(y_1 - y_3) \\
(y_2 + y_4)  +{\rm j}(y_2 - y_4)
\end{array}
\right),
\end{equation} 
and
\begin{equation}
(IV) \hspace{3mm} 
\left(
\begin{array}{c}
x_1+{\rm i}y_1 \\
x_2+{\rm i}y_2 \\
x_3+{\rm i}y_3 \\
x_4+{\rm i}y_4
\end{array}
\right) \hspace{3mm} \leftrightarrow \hspace{3mm}
\frac{1}{\sqrt{2}}\left(
\begin{array}{c}
-(x_2+x_4)  +{\rm j}(x_2-x_4)  \\
(x_1 + x_3) -{\rm j}(x_1-x_3)  \\
-(y_2 + y_4) +{\rm j}(y_2 - y_4) \\
(y_1 + y_3)  -{\rm j}(y_1 - y_3)
\end{array}
\right).
\end{equation}
Four more identifications may be obtained
by a simple `reflection' procedure; simply multiply
each hyperbolic spinor appearing in identifications
(I),(II),(III) and (IV) above by the  
variable ${\rm j}$. This has the effect of interchanging
the `real' and `imaginary' parts of each component
in the spinor.  We have thus enumerated a total of eight 
distinct identifications, and an open question is whether 
there are additional (linearly independent) identifications
that can be made. We leave this question for future work.  
\subsection{Parity}
Under parity, the Dirac 4-spinor $\Psi_{\bf C}$ transforms as follows
\cite{ryder}:
\begin{equation}
 \Psi_{\bf C} \rightarrow
\left(
\begin{array}{cc}
 0 & \mbox{{\bf 1}}_{2 \times 2} \\
 \mbox{{\bf 1}}_{2 \times 2} & 0
\end{array}
\right)
\cdot
\Psi_{\bf C},
\end{equation}
or, in terms of the eight real components $x_i$ and $y_i$,
$i=1,2,3,4$, of the Dirac 4-spinor
$\Psi_{\bf C}$ specified by (\ref{cd}), we have
\begin{equation}
\begin{array}{cccc}
 x_1 \rightarrow x_3 \hspace{5mm}  &
x_2 \rightarrow x_4 \hspace{5mm}  &
x_3 \rightarrow x_1 \hspace{5mm}  &
x_4 \rightarrow x_2 \\
 y_1 \rightarrow y_3 \hspace{5mm}  &
y_2 \rightarrow y_4 \hspace{5mm}  &
y_3 \rightarrow y_1 \hspace{5mm}  &
y_4 \rightarrow y_2 
\end{array}
\end{equation}
According to  the identifications (I),(II),(III) and (IV)
of Section \ref{isomorph},
a parity transformation on $\Psi_{\bf C}$ corresponds 
to ${\bf D}$-conjugation of each component of $\Psi_{\bf D}$. Thus,
$\Psi_{\bf D} \rightarrow \Psi_{\bf D}^{\ast}$ under parity\footnote{
$\Psi_{\bf D}^{\ast}$ denotes taking the ${\bf D}$-conjugate
of each component in $\Psi_{\bf D}$.} for these identifications.
The `reflected' forms of these identifications 
induces the transformation $\Psi_{\bf D} \rightarrow {\rm j}
\Psi_{\bf D}^{\ast}$ under parity. Thus the mathematical
operation of ${\bf D}$-conjugation  
is closely related to the parity symmetry operation.

\section{An Automorphism Algebra of the Dirac Spinor}
\label{auto} 
The existence of distinct equivalences between the transformation
properties of complex (or Dirac)
and hyperbolic spinors
 permits one to construct automorphisms
of the complex Dirac spinor that leave the transformation
properties of its eight real components intact under
Lorentz transformations.

In order to investigate the algebra underlying
the set of all possible automorphisms,
it is convenient to change our current basis to
the so-called `standard representation' of the 
Lorentz group \cite{ryder}. The Dirac 4-spinor
$\Psi^{SR}_{\bf C}$ 
in the standard representation
is related to the original 4-spinor $\Psi_{\bf C}$
according to the relation
\begin{equation}
\Psi^{SR}_{\bf C}
=
\frac{1}{\sqrt{2}}
\left(
\begin{array}{cc}
1 & 1 \\
1 & -1 
\end{array}
\right) \cdot \Psi_{\bf C}.
\end{equation}
The identifications (I)-(IV) stated in Section \ref{isomorph}
are now equivalent to the following identifications:
\begin{equation}
(I)' \hspace{7mm} 
\Psi_{\bf D} =
\left(
\begin{array}{c}
a_1 +{\rm j}b_1  \\
a_2 +{\rm j}b_2  \\
a_3 +{\rm j}b_3 \\
a_4 +{\rm j}b_4
\end{array}
\right) \hspace{5mm} \leftrightarrow \hspace{5mm}
\Psi^{SR}_{\bf C} =
\left(
\begin{array}{c}
a_3+{\rm i}a_1 \\
a_4+{\rm i}a_2 \\
b_3+{\rm i}b_1 \\
b_4+{\rm i}b_2
\end{array}
\right) \label{firstdash}
\end{equation}
\begin{equation}
(II)' \hspace{7mm} 
\Psi_{\bf D} =
\left(
\begin{array}{c}
a_1 +{\rm j}b_1  \\
a_2 +{\rm j}b_2  \\
a_3 +{\rm j}b_3 \\
a_4 +{\rm j}b_4
\end{array}
\right) \hspace{5mm} \leftrightarrow \hspace{5mm}
\Psi^{SR}_{\bf C} =
\left(
\begin{array}{c}
-a_4+{\rm i}a_2 \\
a_3-{\rm i}a_1 \\
b_4-{\rm i}b_2 \\
-b_3+{\rm i}b_1
\end{array}
\right) 
\end{equation}
\begin{equation}
(III)' \hspace{7mm} 
\Psi_{\bf D} =
\left(
\begin{array}{c}
a_1 +{\rm j}b_1  \\
a_2 +{\rm j}b_2  \\
a_3 +{\rm j}b_3 \\
a_4 +{\rm j}b_4
\end{array}
\right) \hspace{5mm} \leftrightarrow \hspace{5mm}
\Psi^{SR}_{\bf C} =
\left(
\begin{array}{c}
-a_1+{\rm i}a_3 \\
-a_2+{\rm i}a_4 \\
-b_1+{\rm i}b_3 \\
-b_2+{\rm i}b_4
\end{array}
\right) 
\end{equation} 
\begin{equation}
(IV)' \hspace{7mm} 
\Psi_{\bf D} =
\left(
\begin{array}{c}
a_1 +{\rm j}b_1  \\
a_2 +{\rm j}b_2  \\
a_3 +{\rm j}b_3 \\
a_4 +{\rm j}b_4
\end{array}
\right) \hspace{5mm} \leftrightarrow \hspace{5mm}
\Psi^{SR}_{\bf C} =
\left(
\begin{array}{c}
a_2+{\rm i}a_4 \\
-a_1-{\rm i}a_3 \\
-b_2-{\rm i}b_4 \\
b_1+{\rm i}b_3
\end{array}
\right).
\end{equation} 
In addition, we have four
more which correspond to the `reflected' form
of the above identifications, and are obtained by interchanging
the `real' and `imaginary' parts of the components of
$\Psi_{\bf D}$:
\begin{equation}
(V)' \hspace{7mm} 
\Psi_{\bf D} =
\left(
\begin{array}{c}
a_1 +{\rm j}b_1  \\
a_2 +{\rm j}b_2  \\
a_3 +{\rm j}b_3 \\
a_4 +{\rm j}b_4
\end{array}
\right) \hspace{5mm} \leftrightarrow \hspace{5mm}
\Psi^{SR}_{\bf C} =
\left(
\begin{array}{c}
b_3+{\rm i}b_1 \\
b_4+{\rm i}b_2 \\
a_3+{\rm i}a_1 \\
a_4+{\rm i}a_2
\end{array}
\right) 
\end{equation}
\begin{equation}
(VI)' \hspace{7mm} 
\Psi_{\bf D} =
\left(
\begin{array}{c}
a_1 +{\rm j}b_1  \\
a_2 +{\rm j}b_2  \\
a_3 +{\rm j}b_3 \\
a_4 +{\rm j}b_4
\end{array}
\right) \hspace{5mm} \leftrightarrow \hspace{5mm}
\Psi^{SR}_{\bf C} =
\left(
\begin{array}{c}
-b_4+{\rm i}b_2 \\
b_3-{\rm i}b_1 \\
a_4-{\rm i}a_2 \\
-a_3+{\rm i}a_1
\end{array}
\right) 
\end{equation}
\begin{equation}
(VII)' \hspace{7mm} 
\Psi_{\bf D} =
\left(
\begin{array}{c}
a_1 +{\rm j}b_1  \\
a_2 +{\rm j}b_2  \\
a_3 +{\rm j}b_3 \\
a_4 +{\rm j}b_4
\end{array}
\right) \hspace{5mm} \leftrightarrow \hspace{5mm}
\Psi^{SR}_{\bf C} =
\left(
\begin{array}{c}
-b_1+{\rm i}b_3 \\
-b_2+{\rm i}b_4 \\
-a_1+{\rm i}a_3 \\
-a_2+{\rm i}a_4
\end{array}
\right) 
\end{equation} 
\begin{equation}
(VIII)' \hspace{7mm} 
\Psi_{\bf D} =
\left(
\begin{array}{c}
a_1 +{\rm j}b_1  \\
a_2 +{\rm j}b_2  \\
a_3 +{\rm j}b_3 \\
a_4 +{\rm j}b_4
\end{array}
\right) \hspace{5mm} \leftrightarrow \hspace{5mm}
\Psi^{SR}_{\bf C} =
\left(
\begin{array}{c}
b_2+{\rm i}b_4 \\
-b_1-{\rm i}b_3 \\
-a_2-{\rm i}a_4 \\
a_1+{\rm i}a_3
\end{array}
\right).
\end{equation} 
Recall what these identifications mean; namely, under any
given Lorentz transformation [Eqn(\ref{transformII})] of
$\Psi_{\bf D}$, the eight real components $a_i$ and $b_i$
($i=1,2,3,4$) transform in exactly the same way as the
eight real components $a_i$ and $b_i$ that appear in
the (eight) complex spinors
$\Psi^{SR}_{\bf C}$ listed
above, after being acted on by the 
corresponding complex Lorentz transformation\footnote{
We assume the $E_i$'s and $F_i$'s are now in the standard representation.} 
[Eqn(\ref{transformD})].

We now define an operator $\rho_{II}$ which takes the complex spinor
$\Psi^{SR}_{\bf C}$ in the identification (I)' above 
and maps it to the complex spinor $\Psi^{SR}_{\bf C}$
in the identification (II)'. Thus $\rho_{II}$ is defined by
\begin{equation}
\rho_{II}\cdot
\left(
\begin{array}{c}
x_1+{\rm i}y_1 \\
x_2+{\rm i}y_2 \\
x_3+{\rm i}y_3 \\
x_4+{\rm i}y_4
\end{array}
\right)
 =
\left(
\begin{array}{c}
-x_2+{\rm i}y_2 \\
x_1-{\rm i}y_1 \\
x_4-{\rm i}y_4 \\
-x_3+{\rm i}y_3
\end{array}
\right),
\end{equation}
for any real variables $x_i$ and $y_i$.
Similarly, we may construct the operators
$\rho_{III}, \rho_{IV}, \dots , \rho_{VIII}$,
whose explicit form we omit for brevity.

If we let 
${\cal V}(\Psi^{SR}_{\bf C})$ denote the eight-dimensional 
vector space 
formed by all {\em real} linear combinations of 
complex 4-spinors, then the linear map $\rho_{II}$,
for example, is
an automorphism of ${\cal V}(\Psi^{SR}_{\bf C})$.
In particular, the transformation properties
of the eight real components of $\Psi^{SR}_{\bf C}$
under a Lorentz transformation is identical to the 
transformation properties of the transformed spinor
$\rho_{II}(\Psi^{SR}_{\bf C})$ under the same Lorentz transformation.
One can show that the set of eight operators
\begin{equation}
 \{1,\rho_{II},\rho_{III},\dots,\rho_{VIII}\} 
\end{equation}
generate an eight dimensional closed algebra with respect to
the real numbers.. 
The subset $\{1,\rho_{II},\rho_{III},\rho_{IV}\}$, for
example, generates the algebra of quaternions. 

One may also consider all commutators of the seven elements
$\rho_{II},\rho_{III},\dots,\rho_{VIII}$. These turn out to
generate a Lie algebra that is isomorphic to 
$\mbox{SU(2)}\times\mbox{SU(2)}\times\mbox{U(1)}$.
The $\mbox{SU(2)}\times\mbox{SU(2)}$ part 
is a Lorentz symmetry. The U(1) factor is
intriguing.

As we pointed out earlier, we have not established that
the algebra generated by the eight operators 
$\{1,\rho_{II},\rho_{III},\dots,\rho_{VIII}\}$ is maximal;
additional independent automorphism operators could exist.
We leave this question for a future investigation. 

\section{Discussion} 
In this work, we constructed a representation of
the six-dimensional Lorentz group in terms of 
$4 \times 4$ generating matrices
defined over the hyperbolic number system, ${\bf D}$.

The transformation properties
of the eight real components of the
corresponding `hyperbolic' 4-spinor under a Lorentz
transformation
was shown to be equivalent to the transformation 
properties of the eight real components
of the familiar complex Dirac spinor, after
making an appropriate identification of components. 
The non-uniqueness
of this identification led to an automorphism algebra 
defined on the vector space of Dirac spinors.
These automorphisms have the property of preserving the
transformation properties of the eight-real components
of a Dirac 4-spinor in any given Lorentz frame. Properties
of this algebra were studied, although we were unable to
prove that the algebra studied here was maximal.

It is interesting to note that the hyperbolic representation
of the Lorentz group turns out to be a subgroup of 
the (fifteen dimensional) special unitary group SU(4,${\bf D}$).
A simple consequence is that $\Psi_{\bf D}^{\dagger} \Psi_{\bf D}$
is a Lorentz invariant scalar. Moreover, after identifying 
$\Psi_{\bf D}$ with $\Psi^{SR}_{\bf C}$, as in equation 
(\ref{firstdash}), for example, it becomes manifest that
the six-dimensional complex representation of the Lorentz
group is a subgroup of SU(2,2;${\bf C}$). This group
is also fifteen dimensional, and it is tempting to assume 
that SU(4,${\bf D}$) and SU(2,2;${\bf C}$) are isomorphic.
This seems to be supported by the proven correspondence
SU$(2,{\bf D}) \cong \mbox{SU}(1,1;{\bf C})$.

However, from general arguments, we were able to assert that
SU$(n,{\bf D})$ and $\mbox{SL}(n,{\bf R})$ 
possess isomorphic Lie algebras for $n \geq 2$. But we also know
SL(4,${\bf R}$) $\cong$ SO(3,3;${\bf R}$) \cite{group}, and
so we conclude that the Lie algebra of SU$(4,{\bf D})$
is isomorphic to the Lie algebra of SO(3,3;${\bf R}$).
But this symmetry evidently
differs from the algebra of generators of the conformal
group SU($2,2;{\bf C}$), which is equivalent to the algebra for 
SO(2,4;${\bf R}$). Thus SU$(4,{\bf D})$ and SU(2,2;${\bf C}$)
are inequivalent groups.

Thus, from the viewpoint of naturally
embedding the Lorentz symmetry into some larger group,
the hyperbolic and complex representations stand apart.
We leave the physics of SU(4,${\bf D}$) as an intriguing 
topic yet to be studied.

 \medskip

\begin{large}{\bf Acknowledgment}
\end{large}
I would like to thank the British Council for support
during the early stages of this work.

\end{document}